\documentclass{elsart}

\usepackage{adnd}
\usepackage{longtable}

\usepackage{mathptmx}

\usepackage{amsmath}

\usepackage{amssymb,amstext}
\usepackage[pdftex,
 letterpaper=true,
 hyperindex=true,
 breaklinks=true,
 colorlinks=false,
 citecolor=blue,
 pdftitle={},
 pdfauthor={}]
{hyperref}

\usepackage{graphicx}

\begin{document}

\begin{frontmatter}

\journal{Atomic Data and Nuclear Data Tables}

\copyrightholder{Elsevier Science}

\runtitle{Cobalt}
\runauthor{Szymanski}


\title{Discovery of the Cobalt Isotopes}


\author{T. Szymanski}
\and
\author{M.~Thoennessen\corauthref{cor}}\corauth[cor]{Corresponding author.}\ead{thoennessen@nscl.msu.edu}

\address{National Superconducting Cyclotron Laboratory and \\ Department of Physics and Astronomy, Michigan State University, \\East Lansing, MI 48824, USA}

\date{July 22, 2009} 

\begin{abstract}
Twenty-six cobalt isotopes have so far been observed; the discovery of these isotopes is discussed. For each isotope a brief summary of the first refereed publication, including the production and identification method, is presented.
\end{abstract}

\end{frontmatter}





\newpage
\tableofcontents
\listofDtables

\vskip5pc

\section{Introduction}\label{s:intro}

 The next paper in the series of the discovery of isotopes \cite{Gin09,Sho09a,Sch09a,Fri09,Hei09,Bur09,Sch09b,Sho09b,Sch09c,Sho09c}, the discovery of the cobalt isotopes is discussed. The purpose of this series is to document and summarize the discovery of the isotopes. Guidelines for assigning credit for discovery are (1) clear identification, either through decay-curves and relationships to other known isotopes, particle or $\gamma$-ray spectra, or unique mass and Z-identification, and (2) publication of the discovery in a refereed journal. The authors and year of the first publication, the laboratory where the isotopes were produced as well as the production and identification methods are discussed. When appropriate, references to conference proceedings, internal reports, and theses are included. When a discovery includes a half-life measurement, we compared the measured value to the most recent adopted value from the NUBASE evaluation \cite{Aud03} or the ENSDF database \cite{ENS08}. In cases where the reported half-life differed significantly from the adopted half-life (up to approximately a factor of two), we searched the subsequent literature for indications that the measurement was erroneous. If that was not the case we credited the authors with the discovery in spite of the inaccurate half-life.

\section{Discovery of $^{50-75}$Co}

Twenty-six cobalt isotopes from A = $50-75$ have been discovered so far; these include one stable, nine proton-rich and 16 neutron-rich isotopes. According to the HFB-14 model \cite{Gor07}, $^{86}$Co should be the last odd-odd particle stable neutron-rich nucleus, and the odd-even particle stable neutron-rich nuclei should continue through $^{91}$Co. $^{50}$Co is the most proton-rich long-lived cobalt isotope because $^{49}$Co has been shown to be unbound \cite{Bla94}. About 14 isotopes have yet to be discovered. Over 60\% of all possible cobalt isotopes have been produced and identified so far. The HFB-14 model was chosen just as an example, and there can be large differences between the different mass models.

Figure \ref{f:year} summarizes the year of first discovery for all cobalt isotopes identified by the method of discovery. The range of isotopes predicted to exist is indicated on the right side of the figure. The radioactive cobalt isotopes were produced using heavy-ion fusion evaporation (FE), light-particle reactions (LP), proton-capture reactions (PC), photo-nuclear reactions {PN), deep-inelastic reactions (DI), and projectile fragmentation or fission (PF). The stable isotope was identified using mass spectroscopy (MS). Heavy ions are all nuclei with an atomic mass larger than A = 4 \cite{Gru77}. Light particles also include neutrons produced by accelerators. In the following paragraphs, the discovery of each cobalt isotope is discussed in detail.

\begin{figure}
	\centering
	\includegraphics[scale=.5]{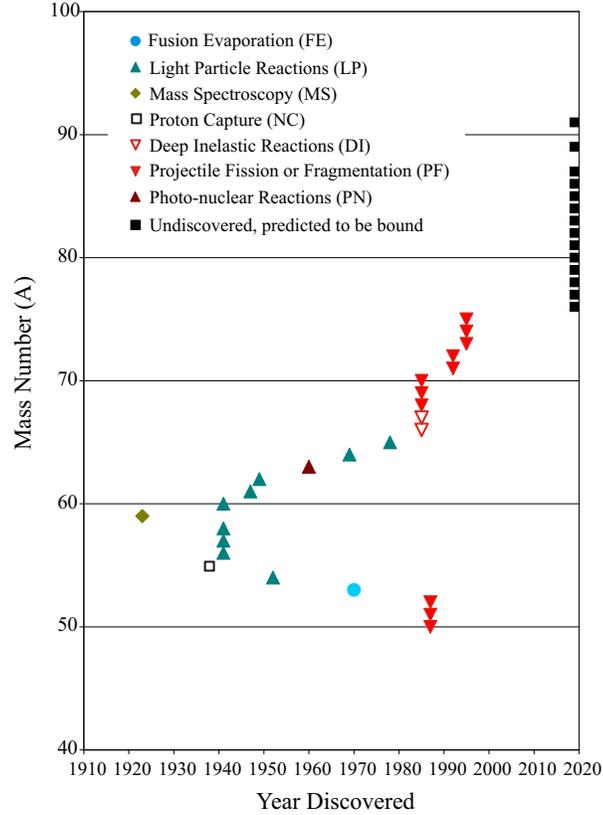}
	\caption{Cobalt isotopes as a function of time they were discovered. The different production methods are indicated. The solid black squares on the right hand side of the plot are isotopes predicted to be bound by the HFB-14 model.}
\label{f:year}
\end{figure}

\subsection*{$^{50-52}$Co}\vspace{-0.85cm}

The 1987 paper \textit{Direct Observation of New Proton Rich Nuclei in the Region 23$\leq$Z$\leq$29 Using A 55A.MeV $^{58}$Ni Beam}, reported the first observation of $^{50}$Co, $^{51}$Co and $^{52}$Co by Pougheon \textit{et al.} \cite{Pou87}. A 55 A$\cdot$MeV $^{58}$Ni was used to bombard a 50 mg/cm$^2$ nickel target at the Grand Acc\'el\'erateur National d'Ions Lourds (GANIL) in Caen, France. The proton-rich fragmentation products were separated with the LISE spectrometer and identified by energy loss, time of flight, and magnetic rigidity measurements. ``The following twelve isotopes were observed for the first time: $^{56}$Cu, $^{52}$Co (T$_z = -1$); $^{55}$Cu, $^{51}$Co, $^{47}$Mn, $^{43}$V (T$_z = 3/2$); $^{52}$Ni, $^{50}$Co, $^{48}$Fe, $^{46}$Mn, $^{44}$Cr (T$_z = -2$) and $^{51}$Ni (T$_z = -5/2$).'' 36, 440 and 3250 events were observed for $^{50}$Co, $^{51}$Co and $^{52}$Co, respectively. The events that would be attributed to $^{49}$Co were judged insufficient to prove its existence.

\subsection*{$^{53}$Co}\vspace{-0.85cm}

Jackson \textit{et al.} discovered $^{53}$Co as described in the 1970 paper \textit{$^{53}$Co$^m$: A Proton-Unstable Isomer} \cite{Jac70}. The Harwell variable energy cyclotron accelerated a $^{16}$O beam to 81 MeV and $^{53}$Co was produced in the fusion-evaporation reaction $^{40}$Ca($^{16}$O,2np). The life-time of delayed protons was measured using a $\Delta$E-E telescope. ``It is our opinion that the best explanation for the origin of the observed activity is the weak proton radioactivity of $^{53}$Co$^m$.'' The ground state of $^{53}$Co was estimated to be bound by 1.6~MeV so the observed proton activity was assigned to the decay of an isomeric state. The measured half-life of 245(20)~ms for this isomer agrees with the presently accepted value of 247(12)~ms. Cerny \textit{et al.} \cite{Cer70} confirmed the interpretation of the results as proton radioactivity in a paper submitted on the same date and published in the same issue immediately following the paper by Jackson \textit{et al.}.

\subsection*{$^{54}$Co}\vspace{-0.85cm}

The discovery of $^{54}$Co was first reported in 1952 by Martin and Breckon in \textit{The New Radioactive Isotopes Vanadium 46, Manganese 50, Cobalt 54} \cite{Mar52}. Protons with energies between 12 and 17 MeV from the McGill University cyclotron bombarded iron targets and $^{54}$Co was produced in the reaction $^{54}$Fe(p,n)$^{54}$Co. Positron activities were displayed on a cathode-ray oscilloscope and photographs of the screen were taken for subsequent graphical analysis. The assignment of $^{54}$Co was based on the threshold energy and the \textit{ft} value. ``One is thus led to assign the 0.40, 0.28, and 0.18 sec. activities to the isotopes V$^{46}$, Mn$^{50}$, and Co$^{54}$, respectively.'' The measured half-life agrees with the presently accepted value of 193.23(14)~ms.

\subsection*{$^{55}$Co}\vspace{-0.85cm}

The first assignment of $^{55}$Co was made by Livingood and Seaborg in the 1938 paper \textit{Long-Lived Radio Cobalt Isotopes} \cite{Liv38}.
$^{55}$Co was produced in the proton-capture reaction $^{54}$Fe(p,$\gamma$) at the Radiation Laboratory of the University of California at Berkeley. ``The bombardment of iron for 1 hour with 100 microamperes of 3.2 MeV protons yields a long-lived activity (as well as the 18-hour Co$^{55}$)...'' The half-life agrees with the presently accepted value of 17.53(3)~h. The activity had previously been observed in the reaction $^{54}$Fe(d,n), however, no mass assignment was made \cite{Liv37}. Also, Darling \textit{et al.} had reported this activity (18.2~h) in the same reaction in an abstract for a conference proceeding, but they also did not assign the half-life to an isotope \cite{Dar37}.

\subsection*{$^{56-58}$Co}\vspace{-0.85cm}

In 1941 the isotopes $^{56}$Co, $^{57}$Co and $^{58}$Co were correctly identified for the first time in the article \textit{Radioactive Isotopes of Cobalt} by Livingood and Seaborg \cite{Liv41}. The isotopes had been produced in several light-particle induced reactions, for example, $^{56}$Fe(d,2n)$^{56}$Co, $^{57}$Fe(d,n)$^{57}$Co, and $^{55}$Mn($\alpha$,n)$^{58}$Co. ``The recent mass spectrographic investigation of the cobalt isotopes by Mitchell, Brown, and Fowler \cite{Mit41}, showing that Co$^{59}$ is the only stable isotope of cobalt, taken together with our recently acquired transmutation data, now makes it possible for us to make isotopic assignments for all the cobalt radioactivities with almost complete certainty.'' The assigned half-lives of 72~d ($^{56}$Co), 270~d ($^{57}$Co) and 72~d ($^{58}$Co) agree with the currently accepted values of 77.23(3)~d, 271.74(6)~d and 70.86(6)~d, respectively. The similarity of the half-lives for $^{56}$Co and $^{58}$Co as well as the incorrectly reported existence of $^{57}$Co \cite{Sam36} resulted in incorrect mass assignments \cite{Liv40,Sea40} or made the assignments not possible. Livingood \textit{et al.} had reported a complex of longer periods of 100$-$200~d \cite{Liv37}. In a subsequent paper, Livingood and Seaborg determined two half-lives of 80~d and 180~d \cite{Liv38}. Half-lives of 1 month and 270~d (Perrier \textit{et al.} \cite{Per38}) and 58~d and 265~d (Barresi and Cacciapuoti \cite{Bar39}) had also been published.

\subsection*{$^{59}$Co}\vspace{-0.85cm}

In 1923 Aston stated the discovery of the only stable cobalt isotope, $^{59}$Co in \textit{Further Determinations of the Constitution of the Elements by the Method of Accelerated Anode Rays} \cite{Ast23}. No details regarding the mass spectroscopic observation of cobalt is given. ``Cobalt also appears to be a simple element of mass number 59, as was to be expected from its atomic weight, which has been determined with great care by a number of observers.'' As mentioned above, $^{57}$Co had erroneously also been reported as being stable \cite{Sam36} and only in 1941 was cobalt established to be a monisotopic element \cite{Mit41}.

\subsection*{$^{60}$Co}\vspace{-0.85cm}

$^{60}$Co was uniquely identified for the first time in 1941 by Livingood and Seaborg in the paper \textit{Radioactive Isotopes of Cobalt} \cite{Liv41}. Deuterons bombarded a cobalt target at the Berkeley cyclotron and decay and absorption curves were measured. The measurements had been published previously \cite{Liv38}, however, no mass assignment was made until 1941: ``The recent mass spectrographic investigation of the cobalt isotopes by Mitchell, Brown, and Fowler \cite{Mit41}, showing that Co$^{59}$ is the only stable isotope of cobalt, taken together with our recently acquired transmutation data, now makes it possible for us to make isotopic assignments for all the cobalt radioactivities with almost complete certainty.'' The quoted half-lives of 5.3~y and 10.7~m  for $^{60}$Co agree with the current adopted values of 5.2713(8)~y and 10.467(6)~m for the ground state and an isomeric state, respectively. The short half-live had previously been reported by several authors. Rotblat \cite{Rot35} did assign a half-life of 20~m to $^{60}$Co which is significantly larger than the correct value. Heyn \cite{Hey37} measured the correct value of 11~m but was not able to make a unique mass assignment because of the (incorrect) reports of the existence of a stable $^{57}$Co isotope \cite{Sam36}. Kikuchi \textit{et al.} \cite{Kik37} and Livingood \textit{et al.} \cite{Liv37} measured a half-life of 10~m and 11~m, respectively, without a mass assignment. Evidence for the long half-life had also been reported previously. A half-life of a year or more was tentatively assigned to $^{60}$Co by Sampson \textit{et al.} \cite{Sam36b}. Risser measured a half-life of 2.0(5)~y \cite{Ris37} and assigned it to $^{58}$Co based on the incorrectly reported existence of $^{57}$Co \cite{Sam36}. Half-lives of 10~y (Livingood and Seaborg \cite{Liv38} and Viktorin \cite{Vik38}), and 4.7~y (Barresi and Cacciapuoti \cite{Bar39}) had been published without mass assignments.

\subsection*{$^{61}$Co}\vspace{-0.85cm}

In 1947 Parmley and Moyer reported the first observation of $^{61}$Co in \textit{Cobalt 61 Radioactivity} \cite{Par47}. Enriched samples of $^{61}$Ni were bombarded by neutrons at the Crocker Laboratory 60-inch cyclotron at Berkeley. Beta-decay curves were measured following chemical separation and a half-life of 1.75(5)~h was observed. ``Chemical extractions showed this activity to belong to cobalt, and most likely to Co$^{61}$.'' This half-life agrees with the presently accepted value of 1.650(5)~h.

\subsection*{$^{62}$Co}\vspace{-0.85cm}

$^{62}$Co was discovered by Parmley \textit{et al.} as reported in the 1949 paper \textit{The Radioactivities of Some High Mass Isotopes of Cobalt} \cite{Par49}. 22~MeV deuterons from the Crocker Laboratory 60-inch cyclotron at Berkeley bombarded a beryllium target to generate a beam of neutrons. $^{62}$Co was then produced in the reaction $^{62}$Ni(n,p) on an enriched $^{62}$Ni target. Following chemical separation, $\beta$-decay curves were measured with a Geiger counter. ``Neutron bombardment of the nickel sample enriched in isotope 62 yielded activities with half-lives of 13.9$\pm$0.2 min. and 1.6$\pm$0.2 min... These facts make it apparent that the 13.9-minutes decay is associated with an (n,p) reaction in Ni$^{62}$ yielding radioactive Co$^{62}$.'' The observed half-lives of 1.6(2)~m and 13.9(2)~m assigned to $^{62}$Co agree with the currently adopted values of 1.50(4)~m and 13.91(5)~m for the ground state and an isomeric state, respectively.

\subsection*{$^{63}$Co}\vspace{-0.85cm}

Morinaga \textit{et al}. discovered $^{63}$Co in 1960 at Tohoku University in Sendai, Japan, which was reported in \textit{Three New Isotopes, $^{63}$Co, $^{75}$As, $^{81}$As} \cite{Mor60}. Nickel was bombarded with 25~MeV bremsstrahlung in the betatron and produced via the reaction $^{64}$Ni($\gamma$,p). ``After a 1.5 minutes bombardment with 25-MeV bremsstrahlung, an activity with a half-life of an order of one minute was observed. Since this activity could not be assigned to any known isotope to be produced by photoreactions on nickel and was too strong for being assigned to any impurity, it was suspected that this might belong to Co$^{63}$.'' Chemical separations were performed to confirm the activity was due to cobalt. Additional measurements at lower energy ruled out $^{62}$Co and the extracted $\beta$-end point energy of 3.6(2)~MeV agreed with $\beta$-decay systematics for $^{63}$Co. The measured half-life of 52(5)~s is significantly larger than the currently accepted value of 26.9(4)~s. Ward \textit{et al.} measured the correct half-life (26(1)~s) in 1969 and suspected that the longer half-life observed by Morinaga \textit{et al.} was due to a contamination from the 10.5~m isomer of $^{60m}$Co produced in the reaction $^{61}$Ni($\gamma$,p) \cite{War69}. We credit Morinaga \textit{et al.} with the discovery of $^{63}$Co because of this explanation for the discrepancy of the half-life and because the correct determination of the $\beta$-endpoint energy. A 28(2)~s half-life had also been observed in 1966 by Strain and Ross, however, they attributed it to $^{64}$Co \cite{Str66}. Also, two months prior to the submission of Morinaga \textit{et al.}, Preiss and Fink had assigned an incorrect half-life of 1.40(5)~h to $^{63}$Co \cite{Pre60}.

\subsection*{$^{64}$Co}\vspace{-0.85cm}

Ward \textit{et al.} assigned a new half-life to $^{64}$Co in the 1969 paper \textit{Decay of $^{63}$Co and $^{64}$Co} \cite{War69}. 14.8 MeV neutrons produced via the $^3$H(d,n)He$^4$ reaction in the University of Arkansas Cockroft-Walton Accelerator bombarded an enriched sample of $^{64}$Ni. $^{64}$Co was then created in the charge-exchange reaction $^{64}$Ni(n,p) and identified by its $\beta$-ray emission. ``The 0.4 sec activity can be tentatively assigned to $^{64}$Co since its $\beta$ end point energy of 7.0$\pm$0.5 MeV is within reasonable agreement with the estimate of 7.5 MeV by Yamada and Matumoto.'' This half-life agrees with the currently adopted value of 0.30(3)~s. Previously, several measurements of different incorrect half-life measurements were reported, for example, 4-5~m by Parmley \textit{et al.} \cite{Par49}, 2.0(2)~m and 7.8(2)~m  for the ground and an isomeric state, respectively by Preiss and Fink \cite{Pre60}, and 28(2)~s by Strain and Ross \cite{Str66}.

\subsection*{$^{65}$Co}\vspace{-0.85cm}

Kouzes and Mueller reported the first observation of $^{65}$Co in the 1978 paper \textit{The Mass of $^{65}$Co} \cite{Kou78}. The Princeton University AVF Cyclotron was used to accelerate $^3$He to 80 MeV which bombarded an enriched $^{70}$Zr target and $^{65}$Co was produced in the five-nucleon pickup reaction $^{70}$Zn($^3$He,$^8$B). The mass of $^{65}$Co was measured with two resistive-wire gas-proportional counters and a plastic scintillator located at the focal plan of a QDDD spectrograph. ``The $^{70}$Zn($^3$He,$^8$B)$^{65}$Co spectrum shows a ground state peak of about 20 nb/sr cross section above background.''

\subsection*{$^{66,67}$Co}\vspace{-0.85cm}

Bosch \textit{et al.} discovered $^{66}$Co and $^{67}$Co in 1985 as described in \textit{Beta-decay half-lives of new neutron-rich chromium-to-nickel isotopes and their consequences for the astrophysical r-process} \cite{Bos85}. $^{76}$Ge was accelerated to 11.4 MeV/u at GSI and bombarded a natural tungsten target. $^{66}$Co and $^{67}$Co were produced in multinucleon transfer reactions and separated with the FEBIAD-F ion source and the GSI on-line mass separator. ``Beta-decay studies of the new neutron-rich isotopes $^{58,59}$Cr, $^{63}$Mn, $^{66,67}$Co and $^{69}$Ni, yielding distincly shorter half-lives than the corresponding theoretical predictions, are presented.'' The measured half-lives of 0.23(2)~s for $^{66}$Co and 0.42(7)~s for $^{67}$Co agree with the currently accepted values of 0.194(17)~s and 0.425(20)~s, respectively.

\subsection*{$^{68-70}$Co}\vspace{-0.85cm}

The 1985 paper \textit{Production and Identification of New Neutron-Rich Fragments from 33 MeV/u $^{86}$Kr Beam in the 18$\leq$Z$\leq$27 Region} by Guillemaud-Mueller \textit{et al.} reported the first observation of $^{68}$Co, $^{69}$Co and $^{70}$Co \cite{Gui85}. The 33 MeV/u $^{86}$Kr beam bombarded tantalum targets and the  fragments were separated with the GANIL triple-focusing analyser LISE. ``Each particle is identified by an event-by-event analysis. The mass A is determined from the total energy and the time of flight, and Z by the $\Delta$E and E measurements.''

\subsection*{$^{71,72}$Co}\vspace{-0.85cm}

In their paper \textit{New neutron-rich isotopes in the scandium-to-nickel region, produced by fragmentation of a 500 MeV/u $^{86}$Kr beam}, Weber \textit{et al.} presented the first observation of $^{71}$Co and $^{72}$Co in 1992 at GSI \cite{Web92}. $^{71}$Co and $^{72}$Co were produced in the fragmentation reaction of a 500 A$\cdot$MeV $^{86}$Kr beam from the heavy-ion synchroton SIS on a beryllium target and separated with the zero-degree spectrometer FRS. ``The isotope identification was based on combining the values of B$\rho$, time of flight (TOF), and energy loss ($\triangle$E) that were measured for each ion passing through the FRS and its associated detector array.'' Forty counts of $^{71}$Co and three counts of $^{72}$Co were recorded. While the identification of $^{71}$Co was unambigious, the assignment of $^{72}$Co was only tentative.

\subsection*{$^{73-75}$Co}\vspace{-0.85cm}

In 1995 Engelmann \textit{et al.} reported the discovery of $^{73}$Co, $^{74}$Co and $^{75}$Co in \textit{Production and Identification of Heavy Ni Isotopes: Evidence for the Doubly Magic Nucleus $^{78}_{28}$Ni} \cite{Eng95}. $^{238}$U ions were accelerated in the UNILAC and the
heavy-ion synchrotron SIS at GSI to an energy of 750 A-MeV. $^{73}$Co, $^{74}$Co and $^{75}$Co were produced by projectile fission, separated in-flight by the FRS and identified event-by-event by measuring magnetic rigidity, energy loss and time of flight. ``For a total dose of 10$^{13}$ U ions delivered in 132 h on the target three events can be assigned to the isotope $^{78}$Ni. Other new nuclei, $^{77}$Ni, $^{73,74,75}$Co and $^{80}$Cu can be identified, the low count rate requires a background-free measurement.'' The observed number of events for $^{73}$Co, $^{74}$Co and $^{75}$Co were 165, 15 and 5, respectively.

\section{Summary}

The identification of the cobalt isotopes close to stability proved to be difficult. The incorrectly reported existence of $^{57}$Co impacted the identification of $^{56-58}$Co and $^{60}$Co. It prevented unique assignments of measured half-lives and resulted in wrong mass assignments. The half-life of $^{55}$Co was first measured without a mass assignment. Prior to the correct identification, a wrong half-life was assigned to $^{63}$Co and the correct half-life was assigned to another isotope. Several wrong half-life measurements were reported for $^{64}$Co and it took twenty years from the first report to the publication of the correct half-life.

\ack

This work was supported by the National Science Foundation under grants No. PHY06-06007.


\newpage

\section*{EXPLANATION OF TABLE}\label{sec.eot}
\addcontentsline{toc}{section}{EXPLANATION OF TABLE}

\renewcommand{\arraystretch}{1.0}

\begin{tabular*}{0.95\textwidth}{@{}@{\extracolsep{\fill}}lp{5.5in}@{}}
\textbf{TABLE I.}
	& \textbf{Discovery of Cobalt Isotopes }\\
\\

Isotope & Cobalt isotope \\
Author & First author of refereed publication \\
Journal & Journal of publication \\
Ref. & Reference \\
Method & Production method used in the discovery: \\
 & FE: fusion evaporation \\
 & LP: light-particle reactions (including neutrons) \\
 & MS: mass spectroscopy \\
 & DI: deep-inelastic reactions \\
 & PC: proton-capture reactions \\
 & PN: photo-nuclear reactions \\
 & PF: projectile fragmentation or projectile fission \\
Laboratory & Laboratory where the experiment was performed\\
Country & Country of laboratory\\
Year & Year of discovery \\
\end{tabular*}
\label{tableI}

\newpage
\datatables

\setlength{\LTleft}{0pt}
\setlength{\LTright}{0pt}


\setlength{\tabcolsep}{0.5\tabcolsep}

\renewcommand{\arraystretch}{1.0}


\begin{longtable}[c]{%
@{}@{\extracolsep{\fill}}r@{\hspace{5\tabcolsep}} llllllll@{}}
\caption[Discovery of Cobalt Isotopes]%
{Discovery of Cobalt isotopes}\\[0pt]
\caption*{\small{See page \pageref{tableI} for Explanation of Tables}}\\
\hline
\\[100pt]
\multicolumn{8}{c}{\textit{This space intentionally left blank}}\\
\endfirsthead
Isotope & Author & Journal & Ref. & Method & Laboratory & Country & Year \\
$^{50}$Co & F. Pougheon & Z. Phys. A & Pou87 & PF & GANIL & France &1987 \\
$^{51}$Co & F. Pougheon & Z. Phys. A & Pou87 & PF & GANIL & France &1987 \\
$^{52}$Co & F. Pougheon & Z. Phys. A & Pou87 & PF & GANIL & France &1987 \\
$^{53}$Co & K.P. Jackson & Phys. Lett. B & Jac70 & FE & Harwell & UK &1970 \\
$^{54}$Co & W.M. Martin & Can. J. Phys. & Mar52 & LP & McGill & Canada &1952 \\
$^{55}$Co & J.J. Livingood & Phys. Rev. & Liv38 & PC & Berkeley & USA &1938 \\
$^{56}$Co & J.J. Livingood & Phys. Rev. & Liv41 & LP & Berkeley & USA &1941 \\
$^{57}$Co & J.J. Livingood & Phys. Rev. & Liv41 & LP & Berkeley & USA &1941 \\
$^{58}$Co & J.J. Livingood & Phys. Rev. & Liv41 & LP & Berkeley & USA &1941 \\
$^{59}$Co & F. W. Aston & Nature & Ast23 & MS & Cavendish & UK &1923 \\
$^{60}$Co & J.J. Livingood & Phys. Rev. & Liv41 & LP & Berkeley & USA &1941 \\
$^{61}$Co & T.J. Parmley & Phys. Rev. & Par47 & LP & Berkeley & USA &1947 \\
$^{62}$Co & T.J. Parmley & Phys. Rev. & Par49 & LP & Berkeley & USA &1949 \\
$^{63}$Co & H. Morinaga & J. Phys. Soc. Japan & Mor60 & PN & Tohoku & Japan &1960 \\
$^{64}$Co & T.E. Ward & J. Inorg. Nucl. Chem. & War69 & LP & Arkansas & USA &1969 \\
$^{65}$Co & R.T. Kouzes& Nucl. Phys. A & Kou78 & LP & Princeton & USA &1978 \\
$^{66}$Co & U. Bosch & Phys. Lett. B & Bos85 & DI & Darmstadt & Germany &1985 \\
$^{67}$Co & U. Bosch & Phys. Lett. B & Bos85 & DI & Darmstadt & Germany &1985 \\
$^{68}$Co & D. Guillemaud-Mueller & Z. Phys. A & Gui85 & PF & GANIL & France &1985 \\
$^{69}$Co & D. Guillemaud-Mueller & Z. Phys. A & Gui85 & PF & GANIL & France &1985 \\
$^{70}$Co & D. Guillemaud-Mueller & Z. Phys. A & Gui85 & PF & GANIL & France &1985 \\
$^{71}$Co & M. Weber & Z. Phys. A & Web92 & PF & Darmstadt & Germany &1992 \\
$^{72}$Co & M. Weber & Z. Phys. A & Web92 & PF & Darmstadt & Germany &1992 \\
$^{73}$Co & Ch. Engelmann & Z. Phys. A & Eng95 & PF & Darmstadt & Germany &1995 \\
$^{74}$Co & Ch. Engelmann & Z. Phys. A & Eng95 & PF & Darmstadt & Germany &1995 \\
$^{75}$Co & Ch. Engelmann & Z. Phys. A & Eng95 & PF & Darmstadt & Germany &1995 \\

\end{longtable}

\newpage


\normalsize

\begin{theDTbibliography}{1956He83}

\bibitem[Ast23]{Ast23t} F.W. Aston, Nature {\bf 112}, 449 (1923)
\bibitem[Bos85]{Bos85t} U. Bosch, W.-D. Schmidt-Ott, P. Tidemand-Petersson, E. Runte, W. Hillebrandt, M. Lechle and, F.-K. Thielemann, R. Kirchner, O. Klepper, E. Roeckl, K. Rykaczewski and, D. Schardt, N. Kaffrell, M. Bernas, Ph. Dessagne, and W. Kurcewicz, Phys. Lett. B {\bf 164}, 22 (1985)
\bibitem[Eng95]{Eng95t} Ch. Engelmann, F. Ameil, P. Armbruster, M. Bernas, S. Czajkowski, Ph. Dessagne, C. Donzaud, H. Geissel, A. Heinz, Z. Janas, C. Kozhuharov, Ch. Mieh\'e, G. M\"unzenberg, M. Pf\"utzner, C. R\"ohl, W. Schwab, C. St\'ephan, K. S\"ummerer, L. Tassan-Got and B. Voss, Z. Phys. A {\bf 352}, 351 (1995)
\bibitem[Gui85]{Gui85t} D. Guillemaud-Mueller, A.C. Mueller, D. Guerreau, F. Pougheon, R. Anne, M. Bernas, J. Galin, J.C. Jacmart, M. Langevin, F. Naulin, E. Quiniou, and C. Detraz, Z. Phys. A {\bf 322}, 415 (1985)
\bibitem[Jac70]{Jac70t} K.P. Jackson, C.U. Cardinal, H.C. Evans, H.A. Jelley, and J. Cerny, Phys. Lett. B {\bf 33}, 281 (1970)
\bibitem[Kou78]{Kou78t} R.T. Kouzes and D. Mueller, Nucl. Phys. A {\bf 307}, 71 (1978)
\bibitem[Liv38]{Liv38t} J.J. Livingood and G.T. Seaborg, Phys. Rev. {\bf 53}, 847 (1938)
\bibitem[Liv41]{Liv41t} J.J. Livingood and G.T. Seaborg, Phys. Rev. {\bf 60}, 913 (1941)
\bibitem[Mar52]{Mar52t} W.M. Martin and S.W. Breckon, Can. J. Phys. {\bf 30}, 643 (1952)
\bibitem[Mor60]{Mor60t} H. Morinaga, T. Kuroyanagi, H. Mitsui, and K. Shoda, J. Phys. Soc. Japan {\bf 15}, 213 (1960)
\bibitem[Par47]{Par47t} T.J. Parmley and B.J. Moyer, Phys. Rev. {\bf 72}, 82 (1947)
\bibitem[Par49]{Par49t} T.J. Parmley, B.J. Moyer, and R.C. Lilly, Phys. Rev. {\bf 75}, 619 (1949)
\bibitem[Pou87]{Pou87t} F. Pougheon, J.C. Jacmart, E. Quiniou, R. Anne, D. Bazin, V. Borrel, J. Galin, D. Guerreau, D. Guillemaud-Mueller, A.C. Mueller, E. Roeckl, M.G. Saint-Laurent, and C. Detraz, Z. Phys. A {\bf 327}, 17 (1987)
\bibitem[War69]{War69t} T.E. Ward, P.H. Pile, and P.K. Kuroda, J. Inorg. Nucl. Chem. {\bf 31}, 2679 (1969)
\bibitem[Web92]{Web92t} M. Weber, C. Donzaud, J.P. Dufour, H. Geissel, A. Grewe, D. Guillemaud-Mueller, H. Keller, M. Lewitowicz, A. Magel, A.C. Mueller, G. M\"unzenberg, F. Nickel, M. Pf\"utzner, A. Piechaczek, M. Pravikoff, E. Roeckl, K. Rykaczewski, M.G. Saint-Laurent, I. Schall, C. St\'ephan, K.\"ummerer, L. Tassan-Got, D.J. Vieira, and B. Voss, Z. Phys. A {\bf 343}, 67 (1992)

\end{theDTbibliography}

\end{document}